\documentclass[lettersize,journal]{IEEEtran}
\usepackage{amsmath,amsfonts}
\usepackage{algorithmic}
\usepackage{array}
\usepackage[caption=false,font=normalsize,labelfont=sf,textfont=sf]{subfig}
\usepackage{textcomp}
\usepackage{stfloats}
\usepackage{cite}
\usepackage{url}
\usepackage{verbatim}
\usepackage{graphicx}
\def\BibTeX{{\rm B\kern-.05em{\sc i\kern-.025em b}\kern-.08em
    T\kern-.1667em\lower.7ex\hbox{E}\kern-.125emX}}
\usepackage{balance}
\usepackage{hyperref}
\usepackage{tabularx}

\begin{document}
\title{CovidTracker: A comprehensive Covid-related social media dataset for NLP tasks}
\author{Richard Plant, Amir Hussain \IEEEmembership{Senior Fellow, IEEE}}


\maketitle

\begin{abstract} The Covid-19 pandemic presented an unprecedented global public health emergency, and concomitantly an unparalleled opportunity to investigate public responses to adverse social conditions. The widespread ability to post messages to social media platforms provided an invaluable outlet for such an outpouring of public sentiment, including not only expressions of social solidarity, but also the spread of misinformation and misconceptions around the effect and potential risks of the pandemic. This archive of message content therefore represents a key resource in understanding public responses to health crises, analysis of which could help to inform public policy interventions to better respond to similar events in future. We present a benchmark database of public social media postings from the United Kingdom related to the Covid-19 pandemic for academic research purposes, along with some initial analysis, including a taxonomy of key themes organised by keyword. This release supports the findings of a research study funded by the Scottish Government Chief Scientists' Office that aims to investigate social sentiment in order to understand the response to public health measures implemented during the pandemic. 
\end{abstract}


\section{Introduction}
\label{sec:introduction}

\IEEEPARstart{T}{he} SARS-Cov-2 novel coronavirus popularly known as Covid-19 has provoked a worldwide response since its designation as a worldwide pandemic by the World Health Organisation in March 2020. As of November 2020, almost 45 million cases had been confirmed, with more than 1.1 million deaths reported \cite{WorldHealthOrganization2020}. While the majority of cases involved mild symptoms including cough, fever, diarrhoea and pain, some affected individuals developed much more acute symptoms, such as pneumonia or kidney failure. Especially vulnerable groups including elderly patients and those suffering multiple co-morbidities proved more susceptible to severe infection, and thus presented a correspondingly higher risk of hospitalisation \cite{sohrabi_world_2020}. 

The virus has been transmitted between humans primarily through respiratory action, with recent studies predicting the rate of transmission between non-immune carriers at 3.77\% \cite{wang_deep_2021}. Public health action concentrated on reducing the spread of the virus through travel travel restrictions, temporary closure of businesses and schools, social distancing guidelines, and the implementation of test-and-trace systems. While there were eventually largely successful efforts to create and distribute vaccine treatments \cite{Bakovic2020, Walsh2020, Sadoff2020, Mazumder2020}, initial responses to the pandemic involved a great deal of prophylactic measures, including widespread filter mask wearing and surface disinfection. 

Much current research has responded to the pandemic by urgently striving to understand and communicate aspects of the social impacts to the public as well as combating pervasive myths \cite{Yang2020, Gozzi2020}, a key role for researchers especially when the public is faced with urgent choices demanding good access to information and the tools to understand potential trade-offs between public health, security, and other social goals \cite{Li2020, Pei2020}.

Public health measures designed to restrict the spread of the virus like social distancing—reducing physical contact between people \cite{Badr2020,VoPham2020}—however led to negative social outcomes, such as the closure of many workplaces \cite{Noguchi2020}, restrictions on personal mobility \cite{Bonaccorsi2020}, and the closure of schools and universities \cite{Bahl2020, Stage2020}. These protective measures themselves had profound effects on many aspects of society, including economic \cite{nicola_socio-economic_2020, deb_economic_2022}, social and psychological \cite{marroquin_mental_2020, buzzi_psycho-social_2020, alradhawi_effects_2020}, and indeed in wider health terms \cite{douglas_mitigating_2020}.

Under these circumstances, a great deal of public discourse has moved to online social networks \cite{Wicke2020, Kousha2020}, presenting an excellent opportunity to gauge the mood and responses of users during a period of exceptional pressure and stress. While this has enabled a large number of people to share their experiences and potentially engage positively with additional sources of social solidarity, we note also the potential for malicious or misinformed commentary, with the additional concerns of increased spread of conspiracy theory, hate speech, and trolling \cite{ferrara_misinformation_2020}.

We present a benchmark dataset drawn from online conversations in the United Kingdom about Covid-19-related topics on social networks, collected during various stages of the pandemic response, between January 2020 and April 2021. It is our hope that open access to this data will spur further research into better informing and understanding public sentiment around the pandemic, as well as helping public health professionals to make informed decisions when implementing.

Our key contributions include:

\begin{itemize}
    \item We make available a large-scale message dataset drawn from both the Twitter and Facebook social networks during the Covid pandemic, consisting of more than 9.4 million individual posts. This set is designed for use in natural language processing applications, and is available through the Edinburgh Napier University research repository\footnote{\url{https://www.napier.ac.uk/research-and-innovation/research-search/outputs/covid-19-uk-social-media-dataset-for-public-health-research}}.
    \item In consultation with an expert panel of public health policy analysts and physicians we developed a topic taxonomy to efficiently determine the broad theme of the textual message content, which we use to annotate each message with a marker of discussion subject using a keyword frequency analysis system. 
    \item We present an indicative set of statistics including the most prevalent discussion topics, as well as some key demographics of the user base. This information could prove useful to public health officials when determining public policy responses to future adverse social and health events.
\end{itemize}

\section{Related Work}
\label{sec:relatedwork}

A number of datasets for natural language processing use extracted from social media activity around the Covid-19 pandemic have previously been published. Among those targeting a much wider geographic lens are Chen et al. \cite{chen_tracking_2020}, which adopts 22 keywords to filter worldwide Twitter messages via the streaming API, resulting in a large multi-lingual unlabelled dataset. Lamsal \cite{RabindraLamsalSchoolofComputerandSystemsSciences2020} used a similar approach, filtering with a list of 90 keywords and hashtags, as well as annotating the resultant dataset with sentiment markers. Dimitrov et al. \cite{Dimitrov2020} extended this approach by adopting a 268 keyword list through which they filtered the existing large-scale TweetsKB dataset \cite{fafalios_tweetskb_2018}. Two sets that we found particularly of note that were not produced from Twitter are the Weibo-COV dataset \cite{Hu2020}, extracted from the social network Weibo, and the Instagram dataset compiled by Zarei et al.\cite{Zarei2020}.

In addition to this, multiple smaller datasets have been produced from Twitter encompassing linked niche topics within the broader ambit of Covid-19, including the effects of Hydroxychloroquine \cite{Mutlu2020}, and various sources of treatment misinformation \cite{Dharawat2020}. The diffusion of fake news through social media sources in multiple languages was the focus of several dataset releases \cite{patwa_fighting_2021, kim_fibvid_2021, yang_checked_2021, hayawi_anti-vax_2022}. 

Since the widespread easing of lockdown measures, we note that a number of new or updated datasets have been published by researchers also aiming for our goal of retrospective analysis. Many of these publications are designed for sentiment analysis or topic modelling purposes, including the COVID-Senti \cite{naseem_covidsenti_2021} and Waheeb et al. \cite{waheeb_topic_2022} datasets. This group also includes some interesting information not previously widely studied, including Cheng et al.'s Government Response Event Dataset \cite{cheng_covid-19_2020}, a collection of public policy announcements by official bodies which includes a manual coding structure for type, scope, and targeted population or group. Similar data is presented by Desvars-Larrive et al. \cite{desvars-larrive_structured_2020}. A comprehensive listing of the textual datasets available, as well as those including other knowledge modalities, can be found in the review conducted by Shuja et al. \cite{shuja_covid-19_2021}.

Studies that utilised key social media Covid-related datasets have tended towards classical NLP applications, notably sentiment analysis \cite{chakraborty_survey_2020}, deployed towards understanding topics such as public reception of contact tracing apps \cite{cresswell_understanding_2021, huang_public_2022}, lockdown procedures \cite{gupta_sentiment_2021, sharma_analyzing_2021}, and the spread and propagation of both accurate and inaccurate information \cite{li_characterizing_2020, chakraborty_sentiment_2020}. A novel approach that we found of particular interest involved the differential comparison of sentiment across national, cultural, and language boundaries \cite{imran_cross-cultural_2020, alsabban_comparing_2021, ghasiya_investigating_2021}.

We note the current scholarship around the ethics of releasing the data of participants in social media research without direct consent, especially \cite{Zimmer2010,Metcalf2016,Williams2017}. However, we conducted a thorough assessment of the privacy risk to individuals posed by our research, and along with complying with social network policies and the relevant sections of the General Data Protection Regulation (GDPR), we believe our research is in the legitimate public interest. We have not shared or published direct tweets or posts by individuals, quotes from individuals or names or locations of users who are not public organisations/entities. We have also striven to comply with best practices for user protection \cite{Franzke2020, ESRC2018}, as well as ensuring that no non-public material is included in our dataset.

\section{Methodology}
\label{sec:methodology}

\subsection{Data collection}

Our solution demanded that we implement a platform agnostic collection system capable of integrating with multiple data sources. In order to achieve this, our primary data store was determined to be a Google Cloud-hosted BigQuery database, into which we would load all records. We therefore designed our collection modules for individual platforms as independent parts of the architecture, capable of being flexibly swapped in and out as required.

\begin{figure}[!t]
    \centering
    \includegraphics[width=0.42\textwidth]{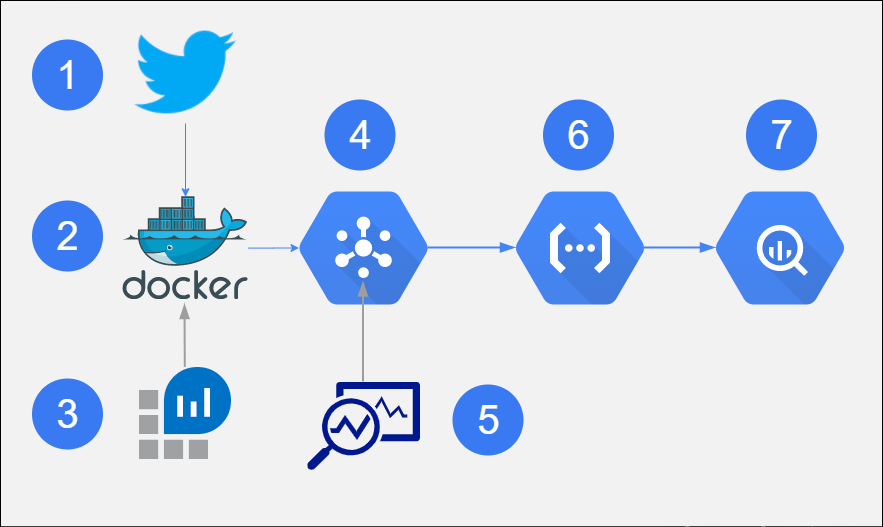}
    \newline
    \caption{Block diagram of Twitter collection pipeline \\
       1. Twitter streaming API 2. Containerised stream listener app \\
       3. Log management 4. Google Pub/Sub subscription \\
       5. App monitoring and alerting 6. Google Cloud Function \\
       7. BigQuery database}
    \label{fig:data_pipeline}
\end{figure}

When working with Twitter data, we designed a containerised streaming listener application \footnote{\url{https://github.com/enu-covid-dashboard/gcp-setup}} that could be deployed quickly on any available host, which would connect to the network endpoint, establish a stream connection with a number of possible pre-selected filters, and consume and log incoming messages constantly. An indicative block diagram of this application can be found in Figure \ref{fig:data_pipeline}.  This application was connected to the endpoint on the 23rd of June at 09:38 GMT, and so the set contains messages from this point onwards.

In this study, we harvested messages within our defined regional boundaries, and defined our stream filter parameters to harvest all messages tagged with a geographical location within the United Kingdom \footnote{Exact bounding box parameters [-6.034129, 49.875160, -0.344459, 61.340529]}. The Twitter 1.1 API specification in use at the time of development \footnote{See \url{https://developer.twitter.com/en/docs/twitter-api/v1/tweets/filter-realtime/api-reference/post-statuses-filter}} allowed the use of bounding box location filters only, and only returns Tweets that have been tagged with Place information derived from the "fine-grained location" permission enabled by users. 

Building on previous work \cite{Fajardo2018,47degrees2018}, we determined the best solution in order to build a low-latency message queuing and ingestion system was via Google Cloud Pub/Sub, through which we sent the extracted post fields to a Cloud Function written in Python that cleaned and sequenced the data into our preferred format, before inserting it into a BigQuery database. See Table \ref{tab:dataformat} for details of the fields included in the imported data.

\begin{table}[ht]
    \centering
    \footnotesize
    \begin{tabular}{|c|c|c|}
        \hline
         \textbf{Platform} & \textbf{Field Name} & \textbf{Data Type} \\
         \hline
         Twitter & Message ID & Integer \\
          & Date Created & Datetime \\
          & Message Text & String \\
          & Location (nearest Place) & String \\
         \hline
         Facebook & Crowdtangle Post ID & String \\
          & Created Date & Datetime \\
          & Message Text & String \\
          & Description Text & String \\
          & Engagement Stats (likes, reacts, etc.) & Integer \\
        \hline
    \end{tabular}
    \vspace*{5mm}
    \caption{Input data formats}
    \label{tab:dataformat}
\end{table}

We gained access to Facebook data through the Crowdtangle platform \footnote{\url{https://www.crowdtangle.com/}}, a Facebook-owned venture which allows access to public post and group data through both visual dashboard tools and an API. In order to harvest Covid-related material, we manually curated lists of important groups, pages and profiles through the web interface. These lists were sorted into both Scotland-only and UK-wide.

A scheduled script \footnote{\url{https://github.com/enu-covid-dashboard/cwd-ingest}} was set up to connect to the API, retrieve all lists attached to the project dashboard, retrieve all posts made by lists members during the preceding 24 hours, then directly upload the results to the BigQuery database. Unfortunately, the lack of a streaming interface precludes near-real-time updates, but this frequency of updates was judged to be acceptable by the project team. Since Facebook posts were available for archive retrieval, we collected posts beginning from the 1st of January 2020 at 00:01 GMT.

We extracted a total of 4,865,020 messages from Twitter between the 23rd of July 2020 and the 12th of March 2021, and a total of 4,564,753 Facebook posts between January 1st 2020 and April 27th 2021. 

\subsection{Processing}

Pre-processing for data was limited to the removal of line ending characters (\textbackslash n and \textbackslash r), as well as annotation with a theme ID determined by keyword frequency analysis, drawn from a pre-determined list of themes and keywords. Any messages that could not have a theme label applied to them were tagged with a valence marker that ensured they would not be entered for further analysis, and have not been published in the released dataset.

\begin{table}[ht]
    \footnotesize
    \centering
    \begin{tabular}{|c|c|c|c|}
        \hline
        \textbf{ID} & \textbf{Theme} & \textbf{Tweets} & \textbf{Facebook posts} \\
        \hline
        1 & Test \& Protect & 30,591 & 40,044 \\  
        2 & Shielding & 605,709 & 563,038 \\  
        3 & Care homes & 14,445 & 38,758 \\  
        4 & Covid survivors & 85,243 & 176,056 \\ 
        5 & Resumption of health services & 24,930 & 84,532 \\
        6 & Mental health \& loneliness & 469,374 & 351,388 \\ 
        7 & Trust in Scottish Government & 40,531 & 24,513 \\  
        8 & Routemap to exit lockdown & 810,092 & 455,360 \\
        9 & Impact on BAME population & 352,224 & 129,883 \\ 
        10 & Inequalities & 84,940 & 63,522 \\ 
        11 & Community cohesion/solidarity & 431,846 & 205,626 \\
        12 & Education & 116,807 & 104,203 \\
        13 & Environment & 26,868 & 35,855 \\
        14 & Quality of life & 36,126 & 74,171 \\
        15 & Social/Family & 417,390 & 522,833 \\
        16 & Leisure/Entertainment & 172,973 & 292,409 \\
        17 & Travel & 52,149 & 78,028 \\
        18 & Business restrictions & 14,276 & 25,317 \\
        19 & Work & 145,113 & 128,619 \\
        20 & Hygiene & 50,405 & 62,532\\
        21 & Shopping & 84,920 & 311,620 \\
        22 & Unemployment & 652,862 & 443,499 \\
        23 & Business growth & 36,630 & 221,613 \\
        24 & Security & 86,303 & 83,642\\
        25 & Other & 22,273 & 47,692 \\
        \hline
        \textbf{Total} & & \textbf{4,865,020} & \textbf{4,564,753} \\
        \hline
    \end{tabular}
    \vspace{5mm}
    \caption{Theme names with associated ID numbers}
    \label{tab:themes}
\end{table}

\begin{figure}[!t]
    \centering
    \includegraphics[width=0.45\textwidth]{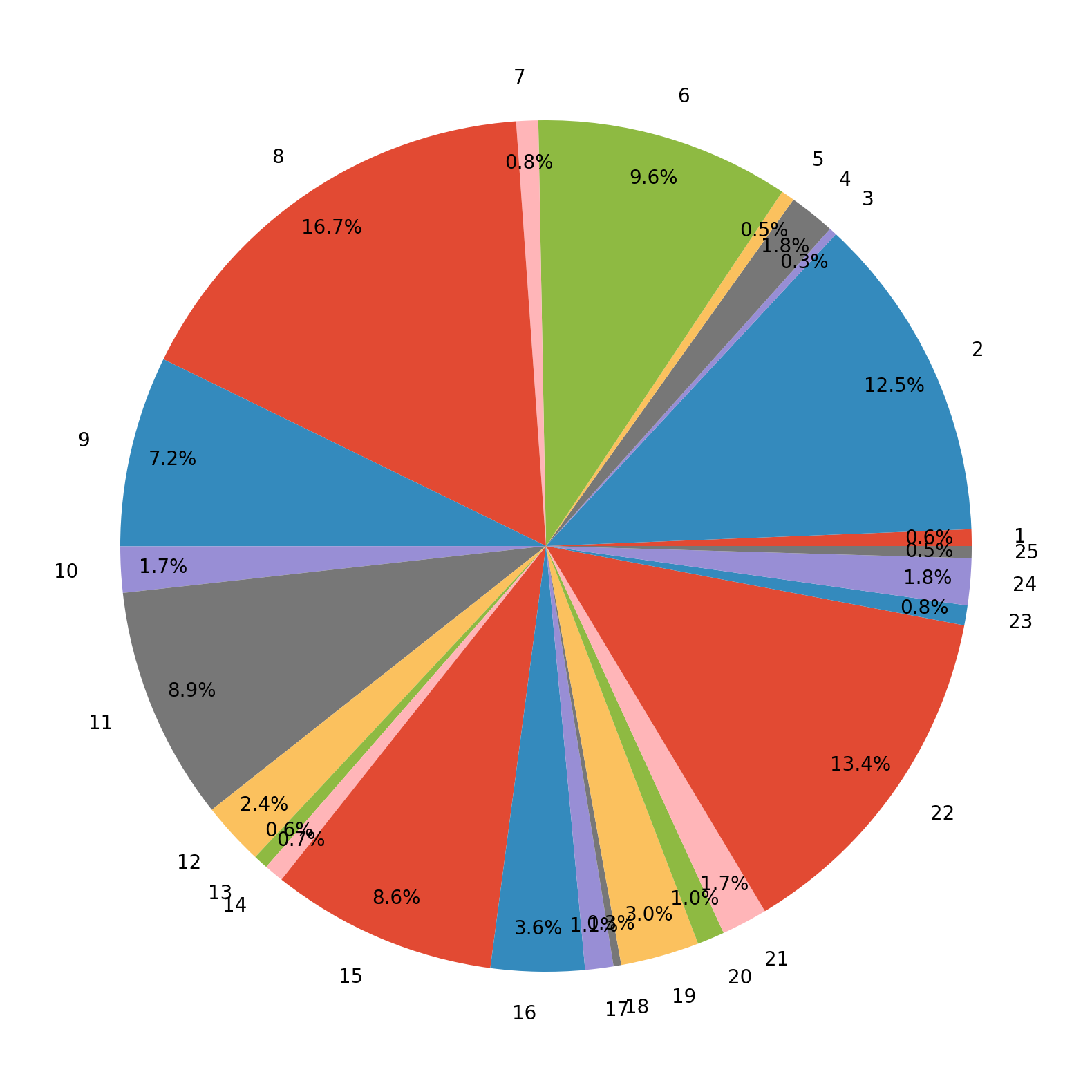}
    \caption{Proportion of each theme in Twitter messages}
    \label{fig:tw_themes}
\end{figure}

\begin{figure}[!t]
    \centering
    \includegraphics[width=0.45\textwidth]{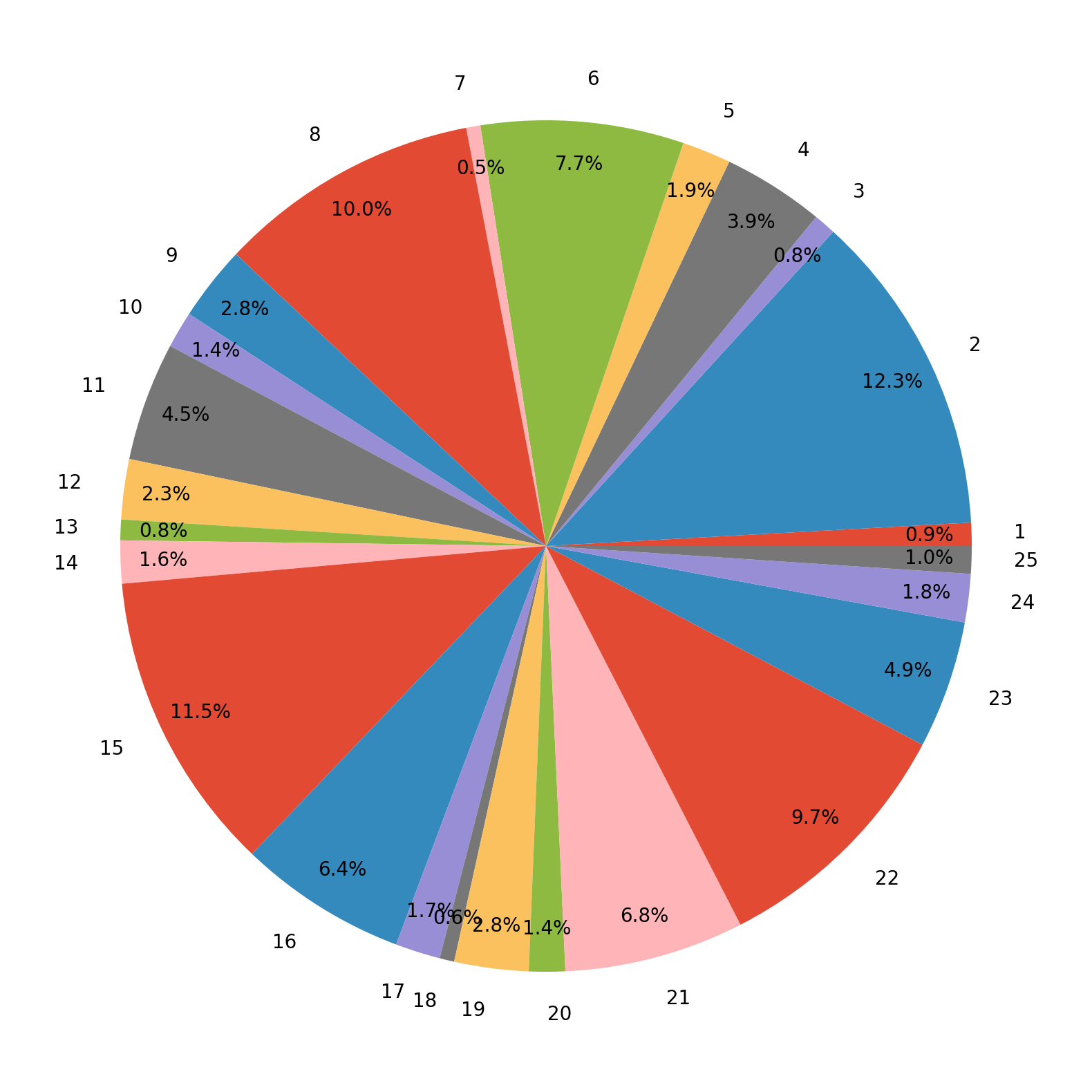}
    \caption{Proportion of each theme in Facebook posts}
    \label{fig:fb_themes}
\end{figure}

A set of priority COVID-19 themes and keywords was developed in close consultation with public health policy domain experts, led by the Covidtracker project co-Lead, Prof Sheikh at Edinburgh University (member of the Scottish Government’s CMO COVID-19 Advisory Group), for potential integration into the Scottish Government framework for decision making. The latter is a policy document setting out the key focus of public health and decision-making bodies when dealing with the pandemic. The COVID-19 themes were also updated in light of public engagement exercises conducted by the government in order to derive priority topics of greatest public interest \cite{govscot_coronavirus_2020}. A full list of the keywords used to filter for each theme can be found in Table \ref{tab:keywords}.

\begin{table}[htbp]
    \scriptsize
    \centering
    \begin{tabularx}{0.5\textwidth}{|c|X|}
        \hline
        \textbf{ID} & \textbf{Keyword/Topic/Idea}  \\
        \hline
        1 & TTI, isolate, test trace, test protect, contact tracing, covid testing, self isolation, 14 day isolation, r number \\ 
        2 & vulnerable, shielding, high risk, elderly, disability, shielded getting out \\ 
        3 & care home, old people home, nursing home, residential care, retirement home \\ 
        4 & survivor, post-covid lethargy, deaths, long covid, long haulers, recover \\ 
        5 & reopen NHS, reopen hospital, reopen GP surgeries, resume dental, reopen dental, non emergency procedure, dentist, health service, routine care, NHS capacity, mammogram, smear test, breast scan, bowel screening, reduction accidents, reduction in viral infections, reduction asthma, blood donations, pregnancy support, non-covid health \\ 
        6 & depression, anxiety, mood, mental health, wellbeing, lonely, loneliness, social isolation, suicide, self harm, insomnia \\ 
        7 & Scotland, Scottish, Scots, Scot Gov, SNP, SG approach, devolved administrations, Nicola Sturgeon, First Minister, Holyrood \\ 
        8 & lockdown, restriction, measures, phase, routemap, mandatory quarantine, guideline, guidance, advice, enforce rules, suppress virus, tackling virus, exit strategy, public compliance, civil liberties, enforcement, freedom, herd immunity, human rights, law, legislation, mass gathering, scientific advice, timeline, circuit breaker \\ 
        9 & racism, ethnicity, minority, ethnicity outcomes, BAME, BME, black, non-white, discrimination, prejudice, disparities, bias, religion, Moslem, Muslim, Islam, Sikh, Hindu, Asian, Indian, Pakistani, Bangladeshi, South Asian, Chinese, Caribbean, Mixed, Multiple, South East Asian, Middle Eastern, Arab, African, Black Caribbean, Black African, Jewish, Jews, ethnic minorities, racial inequality, black ethnic minorities, migrant workforce, xenophobia, hate crime \\ 
        10 & inequalities, rich, well-off, wealthy, working class, homeless, homelessness, poverty, less fortunate, healthy over 70s, older adults, equality, rural communities, digital equality, universal broadband, broadband access, broadband connectivity, vulnerable children, vulnerable families, vulnerable households, vulnerable communities, fair ethical, gender \\ 
        11 & increase xenophobia, crime rate, social stigma, increase racism, community support, increase volunteering, collective solidarity, togetherness, help neighbour, community spirit, rainbows in windows \\
        12 & education, school, home schooling, teaching, blended learning, remote learning, online learning, student, pupil, teachers, virtual classes, class size, nursery, university, additional needs children, special needs child, exam \\ 
        13 & pollution, reduction traffic, green economy, green recovery, waste management, climate impact, environment impact, environment effect, nature impact, nature effect, sustainability, plastic waste, wildlife \\ 
        14 & outdoor exercise, family time, quality of life, life balance, new normal, long term impact, social distancing, stay home \\ 
        15 & visit household, meet household, extended household, gathering, bubble, family visit, family life, social life, social contact, meet friend, see friend, see loved ones, wedding, marriage ceremony, civil partnership, divorce, childcare, places of worship, child adoption, child fostering, domestic abuse, funeral, pets \\ 
        16 & gym, golf, tennis, camping, swimming pool, motorcycling, hill-walking, outdoor activities, indoor activities, running, eating out, exercise, fishing, hospitality, horse riding, personal services, haircut, beauty treatment, facial, massage, cinema \\ 
        17 & drive to exercise, car journeys, holiday plans, holiday abroad, caravan site, visit second home, self-catering, camping, hotel, public transport, travel, tourism, border control, border check, plane, airport, bus, tram, train, air bridge \\ 
        18 & garden centre, barber, hairdresser, beauty salon, recycling centre, property market, construction work, construction site, pub, business damage, cafe, restaurant, shop, retail, shopping centers, nightclub, music venue, theater, concert, business re-opening, body piercing, dog grooming, soft play centre, small business, business restriction, curfew \\ 
        19 & home working, work from home, remote work, office work, key worker, healthcare worker, workplace, employees, early retirement, self employed \\ 
        20 & handwashing, sanitiser, mask, face covering, facial covering, PPE, public hygiene, hygiene standards, hand hygiene \\ 
        21 & cashless, contactless, online shopping, home delivery, supermarkets, shopping trip, shopping habits, high street shops \\ 
        22 & job loss, unemployed, unemployment, made redundant, jobseeker, benefits, furlough, social security, dole, economic uncertainty, basic income, universal credit, deprivation, low income \\ 
        23 & economic recovery, business support, economy, economic impact, business recovery, sector, industry \\
        24 & points system, renew, security risk, service return, statistics, transition, transition arrangements\\
        \hline
    \end{tabularx}
    \vspace{0.5mm}
    \caption{Theme IDs with list of filter keywords}
    \label{tab:keywords}
\end{table}

An indication of how many messages are tagged with each topic ID is displayed in Table \ref{tab:themes}, and as a proportion in Figures \ref{fig:tw_themes} and \ref{fig:fb_themes}. The ratio of themes is similar for each platform, with themes 2,15, and 22 (shielding, social life, unemployment) representing large sections of the discourse, while interesting themes 8, 9, and 11 (ending lockdown, BAME issues, community solidarity) are much more prevalent as a proportion of Twitter messages. This may indicate a differential in the demographic makeup and prevalent concerns of users of the two platforms \cite{gambo_demographics_2020}.

\section{Dataset Features}
\label{sec:features}

\subsection{Publication and Hydration}

In order to comply with network policies for researchers conducting data collection via the Twitter \cite{Twitter} and Crowdtangle \cite{Crowdtangle} platforms, we are able to share only the IDs of material that we collected. This precludes us from sharing the location or text of our collected posts directly.

There are several tools that will enable researchers to rehydrate this data to return the full content of the post or profile. For Twitter, we note that the DocNow Hydrator \footnote{\url{https://github.com/DocNow/hydrator}} and Tweepy Python library \footnote{\url{https://www.tweepy.org/}} can fulfil this function, however the only option for rehydration of Crowdtangle-provided data is to apply for access to the platform and gain access to the official API \footnote{\url{https://github.com/CrowdTangle/API/wiki}}.

Note: there are two separate IDs available for Facebook posts via the Crowdtangle API, the platform ID used by Facebook itself, and the Crowdtangle ID, used by the analytics platform. We have provided the Crowdtangle ID in our dataset, and so when hydrating posts the API endpoint \url{http://api.crowdtangle.com/ctpost/:id} should be used.

Table \ref{tab:dataset_features} shows the fields included in the published dataset, along with a short description of the contents and the data type.

\begin{table}[htbp]
    \centering
    \begin{tabularx}{0.5\textwidth}{|c|c|c|X|}
        \hline
         \textbf{Platform} & \textbf{Field Name} & \textbf{Data Type} & \textbf{Description} \\
         \hline
         Twitter & Date Created & String & Tweet published date\\
          & Message ID & String & Unique Tweet identifier \\
          & Theme ID & Integer & ID of theme allocated by keyword analysis \\
         \hline
         Facebook & Date Created & String & Post published date \\
         & Crowdtangle Post ID & String & Unique identifier for post in Crowdtangle system, in pipe-separated format\\
         & Theme ID & Integer & ID of theme allocated by keyword analysis \\
        \hline
    \end{tabularx}
    \vspace{5mm}
    \caption{Dataset features with data type declaration and description}
    \label{tab:dataset_features}
\end{table}

\subsection{Analysis}

In this section, we carry out some preliminary data analysis using both the Twitter and Facebook datasets as an indication of some potential directions for further research. In conducting this analysis, we randomly sample 10,000 rows without replacement from each set of social network data before rehydrating the message content.

\subsubsection{Word Clouds}

\begin{figure}[!t]
    \centering
    \includegraphics[width=0.4\textwidth]{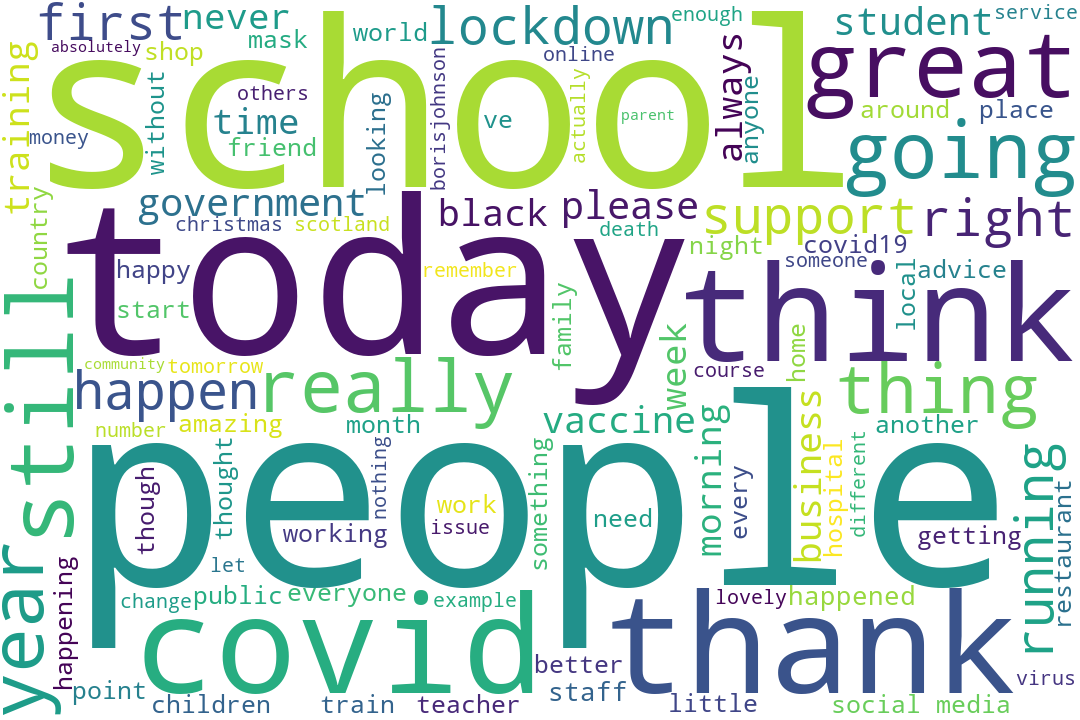}
    \caption{Word cloud for random sampling of Twitter message texts}
    \label{fig:tw_cloud}
\end{figure}

\begin{figure}[!t]
    \centering
    \includegraphics[width=0.4\textwidth]{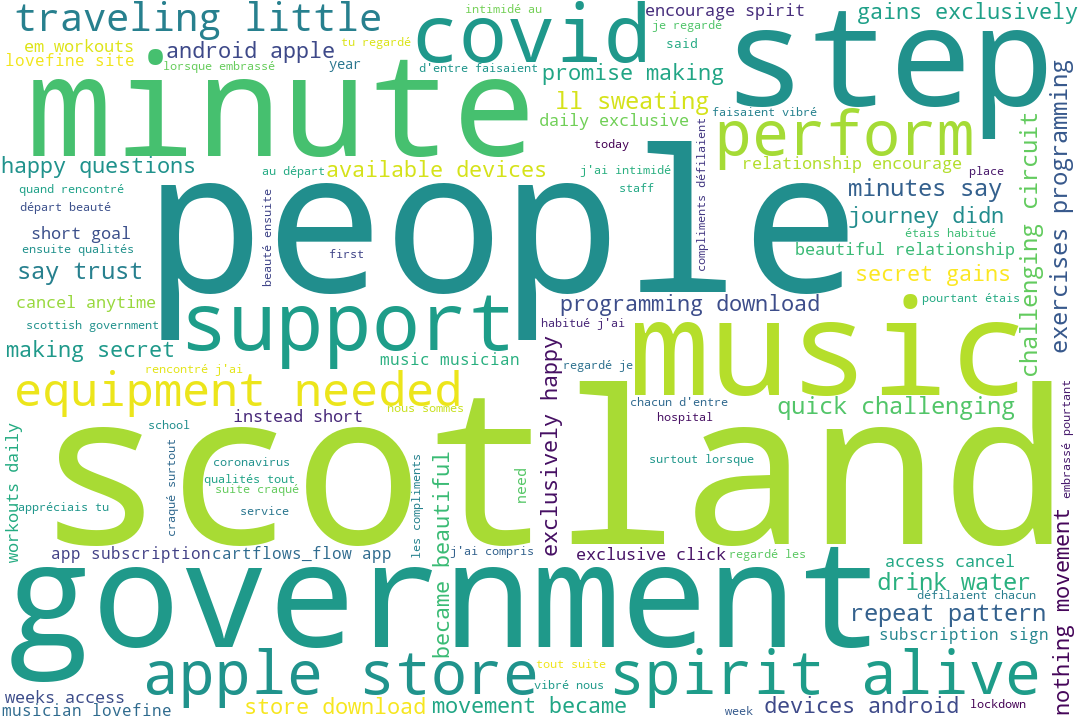}
    \caption{Word cloud for random sampling of Facebook post texts}
    \label{fig:fb_cloud}
\end{figure}

We present here a set of word cloud visualisations of the message content for Twitter (Figure \ref{fig:tw_cloud}) and Facebook (Figure \ref{fig:fb_cloud}) posts. Message texts were first processed to remove stopwords before being ranked by a frequency analysis.

Differences in the topics covered in this small random sample of each network are immediately obvious: Facebook posts appear far more general and less obviously connected to the pandemic, however the key terms do appear with some frequency.

\subsubsection{Topic Modelling}

Using a Latent Dirichlet Allocation (LDA) approach, we model the message texts for each respective platform as the product of a set of unobserved generative topics. Each 'topic' is generally understood to consist of a set of semantically-linked words which taken together define an explanation for the presence of each word in each message text. We define our maximum number of topics as five for the purposes of this analysis. The five most determining words for each topic are presented below in Figures \ref{fig:tw_lda_topics} and \ref{fig:fb_lda_topics}.

Interestingly, while the Twitter topics seem to correlate broadly with Covid-related terms in general discourse at the time, Facebook results vary more widely. Indeed, one topic appears to be made up solely of French terms, perhaps indicating a number of French-speaking ex-patriates or publications among the entries in the dataset. 

\begin{figure}[!t]
    \centering
    \includegraphics[width=0.48\textwidth]{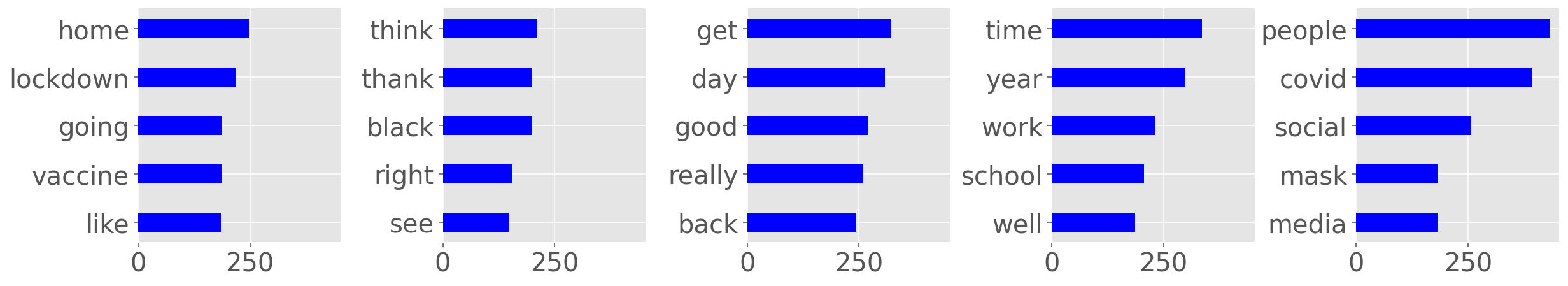}
    \caption{Topics derived from Twitter LDA model}
    \label{fig:tw_lda_topics}
\end{figure}

\begin{figure}[!t]
    \centering
    \includegraphics[width=0.48\textwidth]{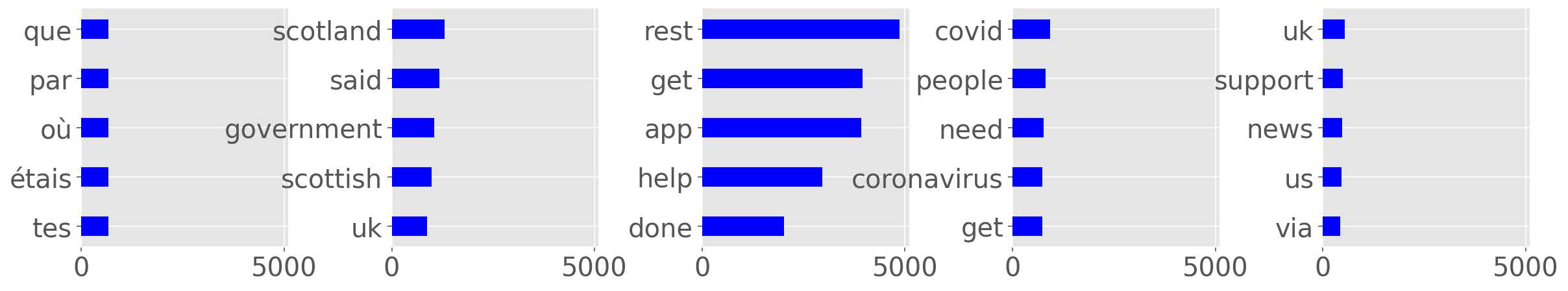}
    \caption{Topics derived from Facebook LDA model}
    \label{fig:fb_lda_topics}
\end{figure}

\section{Conclusion}
\label{sec:conclusion}

In this work, we have presented the outlines of an extensive dataset of social media messages related to the Covid-19 pandemic gathered from users in the United Kingdom from the earliest days of awareness of the disease until after widespread availability of effective vaccines. We have included millions of postings from the two most popular social networks among British users, Twitter and Facebook. Along with providing a rich source of data for multi-modal analyses of public responses to various phases of the crisis, we have also provided a framework for analysing the topics of discussion through keyword analysis conducted with physicians and public policy officials. To facilitate further research, and to promote the goal of further understanding public responses to help engender a more efficient and effective set of public health policies in future, we are releasing this information to the community. 

\section{Further Work}

As detailed in recent research \cite{hussain_opportunities_2021, merchant_public_2021}, more attention should be directed towards social media platforms as important resources for understanding the dissemination and response to public health policy and messaging. In addition to helping drive efforts to combat the 'infodemic' of misinformation that could contribute to negative personal and social decision-making around health topics such as vaccine hesistancy \cite{suarez-lledo_prevalence_2021, banerjee_covid-19_2021}, leveraging social media could lead to increased efficacy of health advocacy and involvement of under-represented populations \cite{jackson_public_2021}.

Previous research which used parts of this dataset has involved both vaccine attitude analysis addressing early-stage (March-November 2020) optimism and concern over vaccine development \cite{hussain_artificial_2021}, and characterization of the nature and frequency of later vaccine mentions in the United Kingdom, in an attempt to better understand vaccine hesitancy trends \cite{hussain_artificial_2022}. These studies used sentiment-based approaches to derive useful findings for public health communications, including that Twitter texts appear to be more likely than Facebook posts to include a negative bias against vaccine uptake, although the absolute ratio of sentiments across the vaccine rollout campaign remained highly positive. In terms of vaccine hesistancy and side effects, the study found that injection-site and short term reactions (pain, redness, swelling, fever, headache) were the most often mentioned, while mentions of blood clots appeared to correlate with press coverage of the AstraZeneca vaccine.

\bibliographystyle{IEEEtran}
\bibliography{references}

\begin{IEEEbiographynophoto}{Richard Plant} is a PhD candidate at Edinburgh Napier University in Scotland. His research interests include statistical privacy tools, ethical machine learning, and robust natural language processing.
\end{IEEEbiographynophoto}

\begin{IEEEbiographynophoto}{Amir Hussain} received the B.Eng. (highest 1st Class Honours with distinction) and PhD degrees from the University of Strathclyde in 1992 and 1997, respectively. Following postdoctoral and academic positions held at the West of Scotland (1996–1998), Dundee (1998–2000), and Stirling Universities (2000–2018), he is currently a Professor and Founding Head of the Cognitive Big Data and Cybersecurity (CogBiD) Research Lab at Edinburgh Napier University. He has (co)authored more than 350 papers, including over a dozen books and around 140 journal papers. Dr. Hussain is Founding Editor-in-Chief of two leading journals: Cognitive Computation (Springer Nature), and BMC Big Data Analytics (BioMed Central), and of the Springer Book Series on Socio-Affective Computing, and Cognitive Computation Trends. He is an Associate Editor for a number of prestigious journals, including Information Fusion (Elsevier), AI Review (Springer), IEEE TRANSACTIONS ON NEURAL NETWORKS AND LEARNING SYSTEMS, IEEE COMPUTATIONAL INTELLIGENCE MAGAZINE, and the IEEE TRANSACTIONS ON EMERGING TOPICS IN COMPUTATIONAL INTELLIGENCE.
\end{IEEEbiographynophoto}

\end{document}